\documentclass[superscriptaddress,preprintnumbers,showpacs,prb,aps,twocolumn]{revtex4}

\usepackage{amsfonts}
\usepackage{amssymb}
\usepackage{amsmath}
\usepackage{graphicx}
\usepackage{natbib}
\usepackage{physics}
\usepackage{siunitx}
\usepackage{float}
\usepackage{xcolor}
\usepackage{siunitx}
\usepackage{array}
\usepackage{tabularx}
\usepackage{multirow}
\usepackage[normalem]{ulem}
\usepackage{array}

\usepackage{fancyhdr}
\setcitestyle{numbers,square}

\usepackage{soul}

\usepackage[hidelinks]{hyperref} 

\begin{document}

\newcommand{\ie}{{\it i.e.}}
\newcommand{\eg}{{\it e.g.}}
\newcommand{\etal}{{\it et al.}}

\newcommand{\micron}{$\mu$m}

\newcommand{\TN}{$T_{\rm N}$}

\newcommand{\Kxx}{$\kappa_{\rm{xx}}$}
\newcommand{\Kxy}{$\kappa_{\rm{xy}}$}
\newcommand{\Kzy}{$\kappa_{\rm{zy}}$}

\newcommand{\cuteo}{Cu$_3$TeO$_6$}

\newcommand{\lco}{La$_2$CuO$_4$}
\newcommand{\nco}{Nd$_2$CuO$_4$}
\newcommand{\scoc}{Sr$_2$CuO$_2$Cl$_2$}



\title{Large Phonon Thermal Hall Conductivity in a Simple Antiferromagnetic Insulator}

\author{Lu~Chen$^{\star}$}
\affiliation{Institut quantique, D\'epartement de physique \& RQMP, Universit\'e de Sherbrooke, Sherbrooke, Qu\'ebec, Canada}

\author{Marie-Eve~Boulanger$^{\star}$}
\affiliation{Institut quantique, D\'epartement de physique \& RQMP, Universit\'e de Sherbrooke, Sherbrooke, Qu\'ebec, Canada}

\author{Zhi-Cheng~Wang}
\affiliation{Department of Physics, Boston College, Boston, MA, USA}

\author{Fazel~Tafti}
\affiliation{Department of Physics, Boston College, Boston, MA, USA}

\author{Louis~Taillefer}
\affiliation{Institut quantique, D\'epartement de physique \& RQMP, Universit\'e de Sherbrooke, Sherbrooke, Qu\'ebec, Canada}
\affiliation{Canadian Institute for Advanced Research (CIFAR), Toronto, Ontario, Canada}

\date{\today}

\begin{abstract}

Phonons are known to generate a thermal Hall effect in certain insulators,
including
oxides with rare-earth impurities,
quantum paraelectrics,
multiferroic materials 
and 
cuprate Mott insulators.
In each case, a special feature of the material is presumed relevant for the underlying mechanism
that confers chirality to phonons in a magnetic field.
The question is whether a phonon Hall effect is an unusual occurrence --
linked to special characteristics such as 
skew scattering off rare-earth impurities,
structural domains,
ferroelectricity, 
ferromagnetism -- 
or a much more common property of insulators than hitherto believed.
To help answer this question,
we have turned to a simple insulator, with none of the previously encountered special features: 
the cubic antiferromagnet \cuteo.
We find that it has the largest thermal Hall conductivity \Kxy~of any insulator so far. 
We show that this record-high \Kxy~signal is due to phonons and
it does not require the presence of magnetic order, as it persists above the 
ordering temperature.
%
%
We conclude that the phonon Hall effect is likely to be a fairly common property of solids. 

%
%
%

\end{abstract}

\maketitle

$^{\star}$ These authors contributed equally to this work.\\

The thermal Hall effect is the thermal analog of the electrical Hall effect. 
Instead of a transverse voltage induced by a perpendicular magnetic field in the presence of an electric current, 
a transverse temperature difference is induced in the presence of a heat current. 
The thermal Hall effect is a consequence of chirality -- a handedness that heat carriers have in a magnetic field. 
Electrons acquire chirality through the Lorentz force acting on charge carriers. 
However, understanding how chirality arises for electrically 
neutral particles -- like phonons, magnons or more exotic excitations -- 
relies on new and mostly unknown mechanisms.

The phonon thermal Hall effect was first observed in the insulator Tb$_3$Ga$_5$O$_{12}$~\cite{Strohm2005,Inyushkin2007},
whose small thermal Hall conductivity \Kxy~was attributed to a special skew scattering of phonons by 
superstoichiometric Tb impurities~\cite{Mori2014}.
Later on, a much larger \Kxy~was measured in the multiferroic material Fe$_2$Mo$_3$O$_8$, a ferrimagnetic insulator,
where it was attributed to phonons in the presence of strong spin-lattice coupling~\cite{Ideue2017}.
More recently, an even larger \Kxy~was reported in two other families of insulators:
the cuprate Mott insulators~\cite{Grissonnanche2019,Boulanger2020,Grissonnanche2020}, such as \lco~and \scoc, 
and the quantum paraelectric SrTiO$_3$~\cite{Li2020}.
There is little doubt that phonons are the bearers of chirality in both families,
but the underlying mechanisms for the Hall effect remain unknown.

The origin of phonon chirality is an open question.
There are two classes of scenarios:
intrinsic scenarios based on the coupling of phonons to their environment
and
extrinsic scenarios based on the skew scattering of phonons by impurities or defects.
For SrTiO$_3$,
the intrinsic effect is the flexoelectric coupling of phonons to their nearly ferroelectric environment,
and the extrinsic factor is the skew scattering of phonons from structural domains~\cite{Chen2020}.
For cuprates,
the intrinsic effect is the coupling of phonons to magnons~\cite{Ye2021}, spinons~\cite{Samajdar2019},
or
a special magnetoelectric order parameter~\cite{Varma2020},
and the extrinsic factors include the scattering of phonons by oxygen vacancies~\cite{Flebus2021},
by pointlike impurities in the presence of a Hall viscosity due to a coupling of phonons to their electronic environment~\cite{Guo2021},
and by resonant skew scatterers~\cite{Sun2021}.

In this Letter, we provide new insights on the origin of phonon chirality by turning to a completely different,
and simple, material: \cuteo.
This is an insulator with a cubic structure, which retains its structure down to low temperature, and therefore does not
harbour structural domains.
It also does not contain rare-earth impurities and is neither a Mott insulator nor a multiferroic or nearly ferroelectric material.
It develops three-dimensional long-range collinear antiferromagnetic order below the N\'eel temperature \TN~$= 63$~K~\cite{Herak2005,Mansson2012}. 

We report the largest thermal Hall conductivity ever observed in an insulator yet,
with \Kxy~$\simeq 1$~W / K m at $T = 20$~K and $H = 15$~T.
This is 50 times larger than in the cuprate \scoc, for example.
Yet the degree of chirality, defined as the ratio $|$\Kxy/\Kxx$|$, where \Kxx~is the longitudinal
thermal conductivity, is the same for these two materials.
This is because the phonon conductivity \Kxx~is 50 times smaller in \scoc, due to
a much stronger scattering by impurities and defects.
The fact that the ratio $|$\Kxy/\Kxx$|$ is the same in these two very different materials,
with different structures, defects and impurities, 
shows that phonon chirality is a much more general phenomenon than hitherto perceived.
Because $|$\Kxy/\Kxx$|$ goes smoothly through \TN,
we infer that antiferromagnetic order {\it per se} is not required,
but we speculate that a coupling of phonons to the spin degrees of freedom may nevertheless play a role.

%
%


\textit{Methods.}---
Single crystals of \cuteo~were grown from CuO powder and TeO$_{2}$ flux.
The starting materials were mixed in a molar ratio of 3:5 and heated to 870$^{\circ}$C at 5$^{\circ}$C/min, 
held for 24 hours, cooled to 700$^{\circ}$C at 1.5$^{\circ}$C/h, and cooled to room temperature at 3$^{\circ}$C/min. 
Crystals of approximate dimensions $4 \times 4 \times 1$~mm$^{3}$
were harvested after washing the solvent with sodium hydroxide and deionized water.
Our sample was
cut and polished in the shape of 
a rectangular platelet,
with the following dimensions (length between contacts $\times$ width $\times$ thickness) : 
$2.2 \times 1.61 \times 0.073$ mm.
%
\cuteo~has a centro-symmetric cubic crystal structure~\cite{Herak2005}.
%
It is not known to undergo any structural transition.
%
The normal to each of the three faces of the sample 
is along each of the three equivalent high-symmetry 
(100) direction of the cubic lattice.
Contacts were made using silver wires and silver paint.
The thermal conductivity \Kxx~and thermal Hall conductivity \Kxy~were measured as described
in refs.~\cite{Grissonnanche2019,Boulanger2020,Grissonnanche2016},
by applying a heat current along the $x$ axis (longest sample dimension)
and a magnetic field along the $z$ axis (normal to the largest face),
and measuring the longitudinal ($\Delta T_{\rm{x}}$)
and transverse ($\Delta T_{\rm{y}}$) temperature differences with type-E 
thermocouples.

\setlength{\tabcolsep}{12pt}

\begin{table}[!]
  \centering
  \label{Table:thermal_conductivity}
  \setlength{\tabcolsep}{8pt}
\begin{tabular*}{\linewidth}[t]{>{\arraybackslash}m{2.3cm} >{\centering\arraybackslash}m{1.0cm} >{\centering\arraybackslash}m{0.55cm} >{\centering\arraybackslash}m{0.65cm} >{\centering\arraybackslash}m{1.2cm}}
    \hline
    \hline
    \noalign{\vskip 0.1cm} 
         \multirow{2}{*}{Material} 
     &$\kappa_{\rm{xy}}$ & $\kappa_{\rm{xx}}$ & $\kappa_{\rm{xy}}/\kappa_{\rm{xx}}$  & $T$, $H$  \\
    \noalign{\vskip 0.05cm}
      & [mW/Km] & [W/Km] & [10$^{-3}$] & [K], [T]  \\
    \noalign{\vskip 0.1cm}
    \hline
    \noalign{\vskip 0.1cm}
Cu$_{3}$TeO$_{6}$  & - 1000 & 330 & - 3.0 & 20, 15 \\
Fe$_2$Mo$_3$O$_8$\cite{Ideue2017}  &12 & 2.5 & 4.8 &  65, 14 \\
Tb$_2$Ti$_2$O$_7$\cite{Hirschberger2015}  & 1.2 & 0.27 & 4.4 & 15, 12  \\
Y$_2$Ti$_2$O$_7$\cite{Hirschberger2015,Li2013}  & 0 & 18 & 0 & 15, 8 \\
\lco\cite{Grissonnanche2019} & - 38  & 12  & - 3.2  & 20, 15  \\
         
\scoc\cite{Boulanger2020} & - 21  & 7 & - 3.0   & 20, 15  \\

\nco\cite{Boulanger2020} & - 200 &  56 & - 3.6 & 20, 15  \\

SrTiO$_3$\cite{Li2020}  &- 80 & 36 & - 2.2 & 20, 12 \\

KTaO$_{3}$\cite{Li2020}  &2 & 32  & 0.06  & 30, 12 \\

RuCl$_3$\cite{Kasahara2018PRL}  &8 & 15.5 & 0.5 & 20, 15  \\
    \noalign{\vskip 0.1cm}
    \hline
    \hline
\end{tabular*}
\caption{Thermal Hall conductivity in various insulators.
The values of \Kxy~and \Kxx~are taken at the specified temperature $T$ and field $H$.
Their ratio gives the degree of chirality.}
\end{table}

\begin{figure}[t]
\centering
\includegraphics[width = 6.2cm]{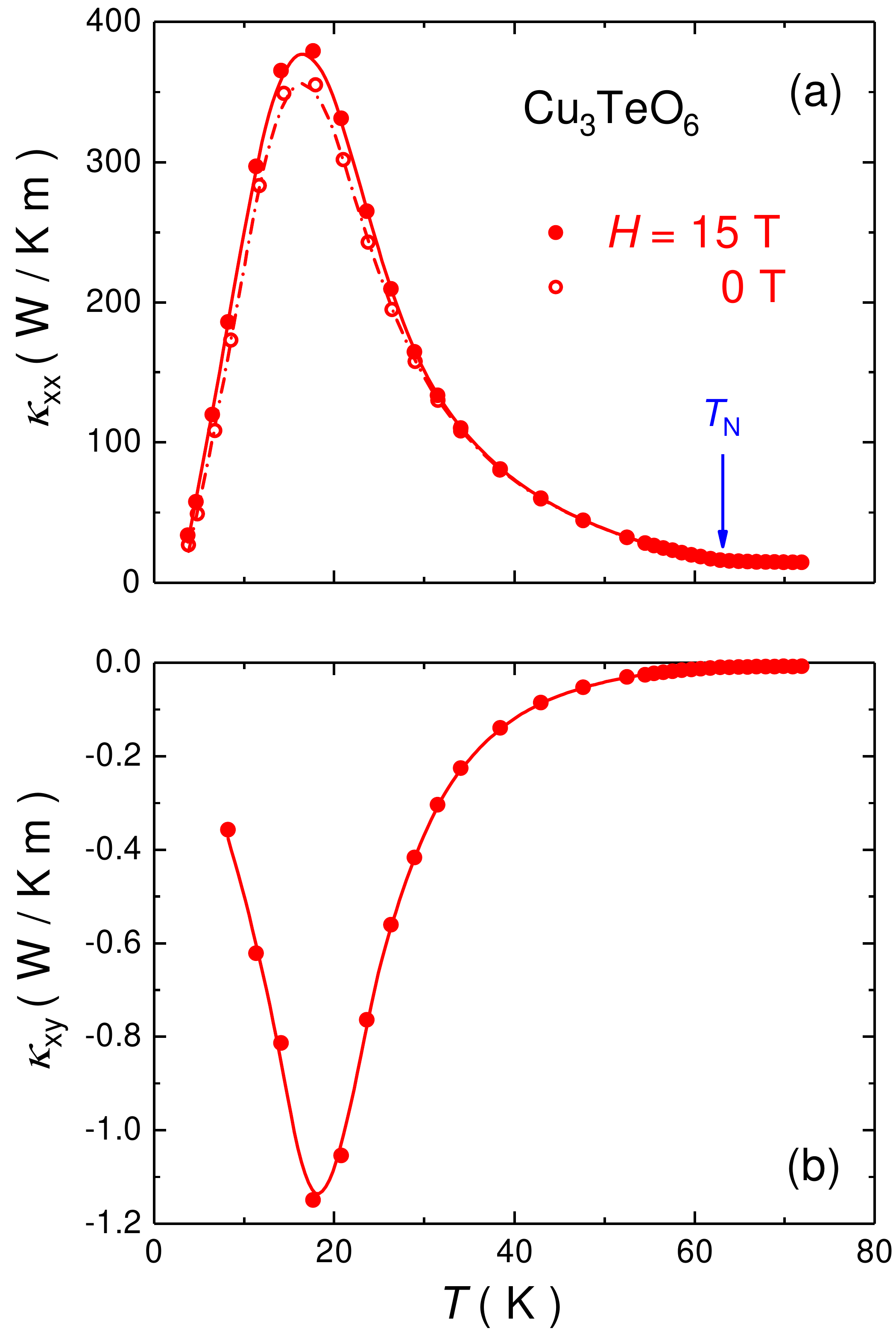}
\caption{(a)
Thermal conductivity \Kxx~of \cuteo~as a function of temperature, 
in zero field (open circles) and in a magnetic field $H = 15$~T (full circles).
The arrow marks the onset of antiferromagnetic order, at \TN~$\simeq 63$~K.
(b)
Corresponding thermal Hall conductivity \Kxy~(at $H = 15$~T). 
Lines are a guide to the eye.
Both \Kxx$(T)$ and \Kxy$(T)$ peak at $T \simeq 17$~K,
following a large increase relative to their values at \TN,
by a factor $\sim 25$ and $\sim 150$, respectively.
The peak value, $|\kappa_{\rm{xy}}| \simeq 1.0$~W/Km, is the largest 
thermal Hall conductivity reported to date in an insulator.
}
\label{fig1}
\end{figure}

\begin{figure}[t]
\centering
\includegraphics[width=6.1cm]{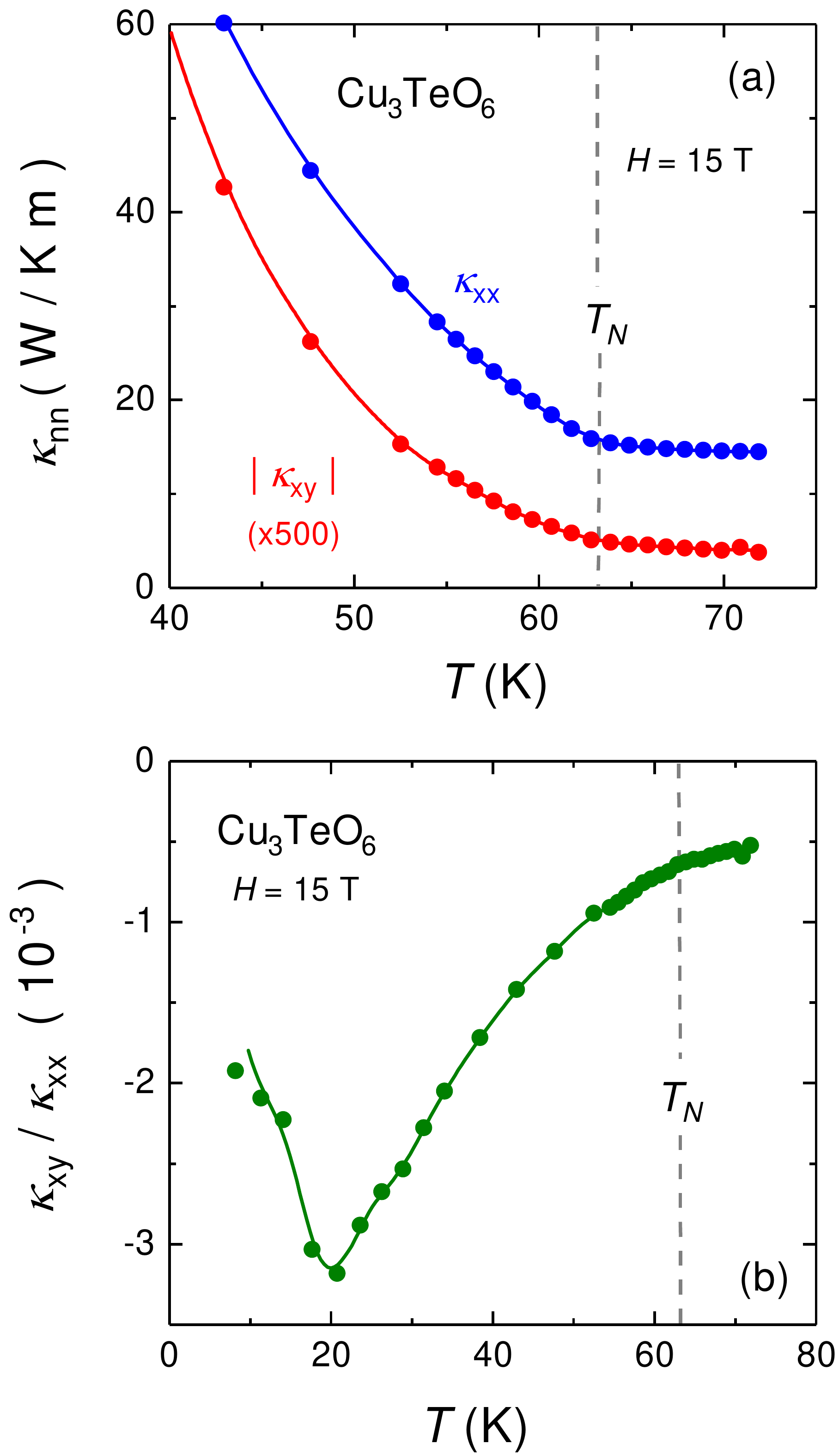}
\caption{
(a)
Comparison of \Kxx$(T)$ (blue) and \Kxy$(T)$ (red; data multiplied by a factor 500) near the antiferromagnetic transition at \TN~(dashed line).
Both curves are seen to rise upon cooling below \TN.
(b)
Ratio of \Kxy$(T)$ over \Kxx$(T)$, vs $T$, at $H = 15$~T.
The magnitude of this ratio increases upon cooling from $T = 70$~K to $T = 20$~K.
The fact that it goes smoothly through \TN~(dashed line) shows that the onset of long-range magnetic order
does not directly affect the thermal Hall effect.
Although \Kxy~in \cuteo~is exceptionally large, the maximal value of the ratio, 
$|\kappa_{\rm xy}/\kappa_{\rm xx}| \simeq 3 \times 10^{-3}$, 
is typical of various insulators (Table~I).
Lines are a guide to the eye.
}
\label{fig2}
\end{figure}


\textit{Thermal conductivity and thermal Hall conductivity.}---
In Fig.~\ref{fig1}(a), the thermal conductivity \Kxx,
measured in a field of 15~T, is plotted as a function of temperature.
%
%
The field dependence of \Kxx~is weak,
being at most 6\% (at $T \simeq 20$~K) and negligible for $T > 30$~K
(Fig.~\ref{fig1}(a)).
Our \Kxx~data are consistent with prior zero-field
data~\cite{Bao2020}.
These authors have argued that although magnons below \TN~can in principle carry heat,
phonons dominate the thermal conductivity of \cuteo,
which is certainly the case above \TN.
\Kxx$(T)$ shows a peak typical of phonons in insulators, located here at $T \simeq 20$~K (Fig.~\ref{fig1}(a)).

In Fig.~\ref{fig2}(a), we zoom on the data near \TN.
Above \TN, \Kxx~is flat,
evidence that phonons are scattered by spin excitations associated with the proximate onset
of antiferromagnetic order~\cite{Bao2020}.
Upon cooling below \TN, \Kxx~suddenly rises, presumably because that scattering is weakened 
when order sets in.

In Fig.~\ref{fig1}(b), the thermal Hall conductivity \Kxy, 
measured on the same sample in the same conditions,
is displayed as a function of temperature.
There is a large negative Hall effect.
We see that \Kxy$(T)$~mirrors the evolution of \Kxx$(T)$,
both peaking at the same temperature.
This immediately suggests that \Kxy~is carried predominently by phonons,
as is \Kxx.
At its peak, $|\kappa_{\rm{xy}}| \simeq 1.0$~W/Km, the largest value of $|\kappa_{\rm{xy}}|$ reported so far in any insulator 
(Table~I).

In Fig.~\ref{fig2}(a), 
we see that \Kxy$(T)$ evolves in tandem with \Kxx$(T)$ across \TN:
it is almost flat above \TN~and rises below \TN. 
This parallel evolution is further
evidence that \Kxy~is carried by phonons.
%
%
It is instructive to plot the ratio \Kxy/\Kxx~vs $T$, as done in Fig.~\ref{fig2}(b),
a quantity that may be viewed as the {\it degree of chirality} -- the extent to which phonons
respond asymmetrically to a magnetic field.
We see that the ratio goes smoothly through \TN, unaltered by the onset of 
antiferromagnetic order.
This shows that long-range order {\it per se} does not play a key role 
in conferring chirality to phonons.
Note that despite the record-high amplitude of $|\kappa_{\rm{xy}}|$ in \cuteo, the ratio \Kxy/\Kxx~is similar
to that found in several other insulators 
(Table~I),
as we discuss below.

\begin{figure*}[t]
\centering
\includegraphics[width=17cm]{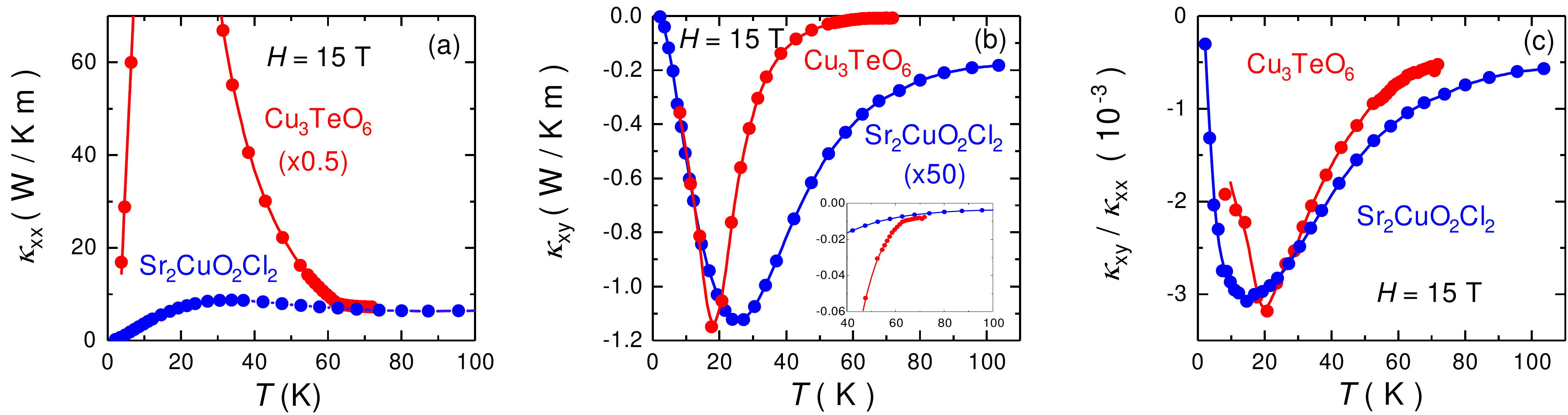}
\caption{
Comparison of two antiferromagnetic insulators, whose thermal transport was
measured in a magnetic field $H = 15$~T:
\cuteo~(red; this work) and the cuprate Mott insulator \scoc~(blue; \cite{Boulanger2020}).
(a)
\Kxx~vs $T$; the data for \cuteo~are multiplied by a factor 0.5. 
(b)
\Kxy~vs $T$; the data for \scoc~are multiplied by a factor 50. 
Inset: zoom between 40~K and 100~K, with no multiplicative factor.
(c)
\Kxy/\Kxx~vs $T$; no multiplicative factor.
All lines are a guide to the eye.
}
\label{fig3}
\end{figure*}


\textit{Discussion.}---
In the antiferromagnetic insulator \cuteo,
two types of neutral excitations can be expected to generate a thermal Hall effect:
magnons and phonons.
We can rule out magnons, based on our empirical evidence and for theoretical reasons.
Empirically,
the fact that 
\Kxy$(T)$ mirrors the temperature evolution of the phonon-dominated \Kxx$(T)$ so well 
(Figs.~\ref{fig1} and~\ref{fig2}(a))
argues against a large contribution to \Kxy~from
magnons. 
Moreover, the fact that the degree of chirality, measured by the ratio \Kxy/\Kxx, 
goes through \TN~seamlessly (Fig.~\ref{fig2}(b)), shows that long-range order, and therefore well-defined magnons,
play little role in \Kxy.
Theoretically, it has been shown that magnons can produce a thermal Hall effect in
antiferromagnetic insulators, but only under certain conditions~\cite{Katsura2010}. 
In a collinear antiferromagnet, a condition is the presence of spin canting due to the Dzyaloshinskii-Moriya (DM) interaction.
Now in \cuteo, theoretical calculations and inelastic neutron scattering experiments show that the collinear antiferromagnetic ground state can be well understood by considering the antiferromagnetic exchange interactions and a global single-ion anisotropy term without introducing any DM interaction~\cite{Yao2018}. 
Neutron powder diffraction results indicate that the possible non-collinear canting of spins is 
no more than 6$^{\circ}$~\cite{Herak2005}. 
Under such conditions, the \Kxy~signal expected from magnons is estimated to be much smaller than the signal
reported in \lco~\cite{Samajdar2019}, which is in turn much smaller than what we observe in \cuteo. 
So, the record-high thermal Hall conductivity in \cuteo~is not generated by magnons.

The only type of heat carriers left that could generate a thermal Hall effect in \cuteo~are phonons. 
Two empirical observations confirm that it is indeed the phonons that generate the huge \Kxy~in \cuteo.
First, \Kxy$(T)$ and \Kxx$(T)$ evolve in parallel across the antiferromagnetic transition, 
both increasing in tandem upon cooling below \TN~$\simeq 63$~K (Fig.~\ref{fig2}(a)).
Second, while the degree of chirality in \cuteo~is not exceptionally high,
what is exceptionally high amongst the insulators for which
a \Kxy~signal has been 
reported is the phonon-dominated \Kxx~(Table~I).
And this is why \Kxy~is so large -- because phonons conduct better.
So, the record-high thermal Hall conductivity in \cuteo~is a property of phonons. 

%

The remaining question
is: 
what makes phonons chiral in \cuteo? 
Two features of our data argue against impurity scattering playing a key role,
and hence favour an intrinsic scenario.
The first is the major change in the dominant scattering process with temperature.
At $T \simeq 20$~K,
where scattering is dominated by impurities, phonon conduction peaks at
\Kxx~$\simeq 350$~W/Km (Fig.~1(a)), a high value that reflects the high quality of our 
single crystal.
By $T \simeq 70$~K, where scattering is dominated by spin fluctuations and other phonons,
\Kxx~has dropped by a factor 25 (Fig.~1(a) and Fig.~2(a)).
Yet despite this complete change of regime, the chirality does not change much between 20~K and 70~K (Fig.~2(b)).
The ratio \Kxy/\Kxx~does decrease (by a factor 6), but such a decrease is a general property of phonon chirality,
unrelated to a change in the dominant scattering process.
This can be seen by comparing to the cuprate Mott insulator \scoc~\cite{Boulanger2020},
an antiferromagnet with \TN~$=250$~K.

In Fig.~\ref{fig3}(a), 
we see that \Kxx~in \scoc~is essentially constant between 20~K and 70~K,
showing that impurity scattering in that material remains the dominant scattering.
Yet \Kxy~drops by a factor 5-6 (Fig.~\ref{fig3}(b)), so that \Kxy/\Kxx~drops by the same factor as in \cuteo~(Fig.~\ref{fig3}(c)), 
which shows
that such a decrease with temperature is a generic property of chiral phonons and 
not the result of a change in the scattering mechanism.
This comparison therefore argues for an intrinsic mechanism in these two materials.

A second argument against an extrinsic scenario is the simple fact that
these two materials have the same degree of chirality even though they have very different levels
of impurity scattering. 
In particular at $T=20$~K, where impurity scattering dominates,
\Kxy/\Kxx~has the same value (Fig.~\ref{fig3}(c)) even though 
impurity scattering is two orders of magnitude stronger in \scoc.
Indeed, the phonon conductivity \Kxx~is lower in \scoc~by a factor 50 (Table~I).
In fact, more generally, the degree of chirality is also the same in other insulators, 
with $|$\Kxy/\Kxx$| = 3-5 \times 10^{-3}$ at $H=15$~T in Fe$_2$Mo$_3$O$_8$~\cite{Ideue2017}
and Tb$_2$Ti$_2$O$_7$~\cite{Hirschberger2015}, even though \Kxx~is two and three orders of magnitude smaller, 
respectively 
(Table~I).

It was recently proposed theoretically that in oxide insulators the skew scattering of phonons 
by charged impurities like oxygen vacancies produces a thermal Hall effect~\cite{Flebus2021}.
However,
experimental data 
suggest
that this mechanism yields a rather small \Kxy~signal.
Indeed, in
the isostructural pyrochlore oxides Tb$_2$Ti$_2$O$_7$ and Y$_2$Ti$_2$O$_7$,
with nominally similar levels of oxygen vacancies, one finds dramatically different Hall responses (Table~I),
with \Kxy~$=0$ in Y$_2$Ti$_2$O$_7$~\cite{Hirschberger2015}.
A huge difference is also found in the closely related oxides SrTiO$_3$ and KTaO$_3$,
where the latter has a very small \Kxy~signal (Table~I)~\cite{Li2020}.
%
These two pairwise comparisons suggest that 
oxygen vacancies
are not responsible for the large \Kxy~signals seen in oxide insulators.
%
(Superstoichiometric rare-earth ions and structural domains could play an extrinsic role
in Tb$_2$Ti$_2$O$_7$ and SrTiO$_3$, respectively, but neither of these mechanisms
applies to \cuteo~or \scoc.)

We conclude that the chirality of phonons detected in 
\cuteo~and \scoc, and most likely several other insulators,
comes from an intrinsic coupling to their environment.
The question is what is that coupling?
Why is it present in some materials and absent in others?
This remains an open question, 
but we speculate that a promising avenue of investigation is the coupling of phonons to spins,
for the simple empirical reason that we are unaware of any magnetic insulator where \Kxy~$=0$.
Several authors have shown theoretically that a phonon thermal Hall effect can arise from 
such a coupling~\cite{Ye2021,Sheng2006,Kagan2008,Wang2009,Zhang2010, Qin2012}.
The next step is to explain the large degree of chirality, with $|$\Kxy/\Kxx$| = 3-5 \times 10^{-3}$ (at $H=15$~T).

\textit{Summary and outlook.}---
We have measured the thermal conductivity \Kxx~and the thermal Hall conductivity \Kxy~of the antiferromagnetic insulator \cuteo.
We report the largest value of $|$\Kxy$|$ ever observed in an insulator.
We provide empirical and theoretical arguments for why \Kxy~must be due to phonons,
primarily.
On the basis of a comparison with the cuprate material \scoc,
which exhibits the same ratio \Kxy/\Kxx, or degree of chirality, both in magnitude and in temperature dependence,
even though its phonon conductivity is 50 times smaller, 
we conclude that the mechanism for phonon chirality is not primarily extrinsic,
{\it i.e.} controlled by impurity or defect scattering.
We  infer that phonons become chiral by virtue of their intrinsic coupling to their environment.
Although the nature of this coupling remains unclear, we propose that a likely possibility is spin-phonon coupling.
Our findings suggest that a large phonon thermal Hall effect may be a common occurrence in magnetic insulators.
This puts two prior studies of the thermal Hall effect into perspective.
First, it raises the question of whether the \Kxy~signal measured in the Kitaev material $\alpha$-RuCl$_{3}$~\cite{Kasahara2018PRL},
hitherto attributed entirely to Majorana fermions~\cite{Kasahara2018}, may in part be due to phonons.
Secondly, it raises the possibility that the phonon thermal Hall conductivity in cuprates that appears upon entering the 
pseudogap phase~\cite{Grissonnanche2019,Grissonnanche2020}, when the doping is reduced below the pseudogap critical point, 
may be the signature of 
antiferromagnetic correlations.

\textit{Acknowledgements.}---
L.T. acknowledges support from the 
Canadian Institute for Advanced Research (CIFAR) as a CIFAR Fellow 
and funding from 
the Institut Quantique, 
the Natural Sciences and Engineering Research Council of Canada (NSERC; PIN:123817), 
the Fonds de Recherche du Qu\'{e}bec -- Nature et Technologies (FRQNT),
the Canada Foundation for Innovation (CFI),
and a Canada Research Chair.
F.T. acknowledges support from the National Science Foundation under award No. DMR-1708929.
This research was undertaken thanks in part to funding from the Canada First Research Excellence Fund.

%

\vfill

\bibliography{reference}

\end{document}